\documentclass[reprint,amsmath,amssymb,aps,pra]{revtex4-1}
\usepackage{amsmath,epsfig,wrapfig,graphicx,color,bm}

\begin{document}
\title{Magnetic phase diagram of the linear quantum ferro-antiferromagnet Cs$_{2}$Cu$_{2}$Mo$_{3}$O$_{12}$}

\author{D.~Flavi{\'a}n}
\email[Masters project: ]{daniefla@student.ethz.ch}
\affiliation{Laboratory for Solid State Physics, ETH Z{\"u}rich, 8093 Z{\"u}rich, Switzerland}
\author{S.~Hayashida}
\email[Corresponding author: ]{shoheih@phys.ethz.ch}
\affiliation{Laboratory for Solid State Physics, ETH Z{\"u}rich, 8093 Z{\"u}rich, Switzerland}
\author{L.~Huberich}
\affiliation{Laboratory for Solid State Physics, ETH Z{\"u}rich, 8093 Z{\"u}rich, Switzerland}
\author{D.~Blosser}
\affiliation{Laboratory for Solid State Physics, ETH Z{\"u}rich, 8093 Z{\"u}rich, Switzerland}
\author{K.~Yu.~Povarov}
\affiliation{Laboratory for Solid State Physics, ETH Z{\"u}rich, 8093 Z{\"u}rich, Switzerland}
\author{Z.~Yan}
\affiliation{Laboratory for Solid State Physics, ETH Z{\"u}rich, 8093 Z{\"u}rich, Switzerland}
\author{S.~Gvasaliya}
\affiliation{Laboratory for Solid State Physics, ETH Z{\"u}rich, 8093 Z{\"u}rich, Switzerland}
\author{A.~Zheludev}
\email{zhelud@ethz.ch}\homepage{http://www.neutron.ethz.ch/}
\affiliation{Laboratory for Solid State Physics, ETH Z{\"u}rich, 8093 Z{\"u}rich, Switzerland}

\date{\today}

\begin{abstract}
A single-crystal sample of the frustrated quasi-one-dimensional quantum magnet Cs$_{2}$Cu$_{2}$Mo$_{3}$O$_{12}$ is investigated by magnetic and thermodynamic  measurements. 
A combination of specific heat and magnetic torque measurements maps out the entire $H$-$T$ phase diagram for three orientations.
Remarkably, a new phase emerges below the saturation field, irrespective of the crystal orientation. 
It is suggested that the presaturation phase represents spin-nematic order or other multi-magnon condensate. 
The phase diagrams within the long-range-ordered dome are qualitatively different for each geometry.
In particular, multiple transitions are identified in the field along the chain direction.
\end{abstract}
 
\maketitle
\section{Introduction}
Quantum magnets with competing antiferromagnetic (AFM) and ferromagnetic (FM) bonds may demonstrate very peculiar behavior, especially in applied magnetic fields. 
This is true for even the simplest $S=1/2$ chain model with  $J_{1}<0$ (FM) nearest-neighbor and $J_2>0$ (AFM) next-nearest-neighbor exchange constants~\cite{Tonegawa1989,Igarashi1989,Tonegawa1990,Chubukov1991}.  
The FM coupling  gives rise to an attraction between magnons in the high-field fully polarized phase, and sometimes stabilizes multi-magnon bound states in that regime~\cite{Meisner2006,Kecke2007,Vekua2007,Hikihara2008,Sudan2009,Zhitomirsky2010}. 
As the magnetic field is lowered, such bound states may condense before a conventional AFM ordering occurs via single-magnon Bose-Einstein condensate (BEC)~\cite{Batyev1984}. 
The result is an entire zoo of possible $n$-magnon condensates. 
These presaturation phases are characterized by exotic symmetry breaking patterns giving rise to unconventional states such as spin-nematic (quadrupolar), octupolar or hexadecapolar order~\cite{Meisner2006,Kecke2007,Vekua2007,Hikihara2008,Sudan2009,Meisner2009,Zhitomirsky2010,Sato2013}, spin density waves (SDW)~\cite{Hikihara2008,Sudan2009,Sato2013,Nishimoto2015}, as well as various chiral structures~\cite{Kolezhuk2005,Hikihara2008,Sudan2009,Meisner2009}. 
The challenge is to find and explore these exotic states of spin matter experimentally.
To date, only a handful have been found in real FM-AFM spin-chain materials: LiCuVO$_{4}$~\cite{Svistov2011,Masuda2011,Mourigal2012,Nawa2013,Buttgen2014,Orlova2017,Gen2019}, PbCuSO$_{4}$(OH)$_{2}$~\cite{Willenberg2012,Schapers2013,Willenberg2016,Feng2018,Heinze2019}, NaCuMoO$_{4}$(OH)~\cite{Nawa2014,Nawa2017,Asai2020}, and $\beta$-TeVO$_{4}$~\cite{Pregelj2015,Weickert2016,Pregelj2019}.

The focus of the present study is on another candidate species, namely the linear chain molybdate Cs$_{2}$Cu$_{2}$Mo$_{3}$O$_{12}$~\cite{Solodovnikov1997,Hase2005}.
This rather complex system has been extensively studied, but only using powder samples. 
Previous work includes thermodynamic and magnetic measurements~\cite{Hase2005,Fujimura2016}, high pressure studies~\cite{Hamasaki2007}, NMR~\cite{Hoshino2014,Goto2017,Yagi2018,Matsui2018}, and muon spin relaxation measurements~\cite{Goto2017,Matsui2018}.
The material orders magnetically in three dimensions at $T_{\rm N} = 1.85$~K~\cite{Fujimura2016}, but its low dimensionality and high degree of frustration are beyond doubt. 
The exchange constants have been estimated as $J_{1}=-93$~K and $J_{2}=33$~K $(J_{2}/|J_{1}|=0.36)$~\cite{Hase2005}.
Particularly intriguing are NMR results which seem to indicate a nematic Tomonaga-Luttinger liquid (TLL) state at high magnetic field~\cite{Hoshino2014,Goto2017}.

Further progress can only be achieved using single-crystal experiments. 
Recent studies of this type were key to understanding the rich physics of  the isostructural Rb$_{2}$Cu$_{2}$Mo$_{3}$O$_{12}$~\cite{Hayashida2019}. 
In the present work, we report comprehensive single-crystal magneto-thermodynamic measurements on Cs$_{2}$Cu$_{2}$Mo$_{3}$O$_{12}$. 
We map out the entire $H$-$T$ phase diagram down to 0.1~K and up to the full saturation for three field directions.
The Cs compound appears to be considerably more exotic than its Rb counterpart.
The most intriguing finding is a distinctive presaturation phase in all field geometries. 
It has all characteristics of a multi-magnon condensate. 
Another is an entire cascade of magnetic phase transitions in a magnetic field applied along the chain axis.

\begin{figure}[tbp]
\includegraphics[scale=1]{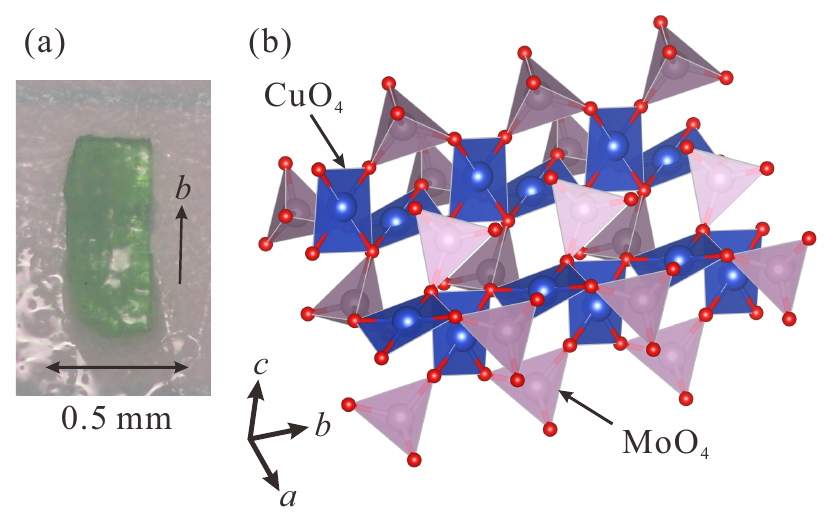}
\caption{(a) A single-crystal Cs$_{2}$Cu$_{2}$Mo$_{3}$O$_{12}$ sample used in this work. (b) Schematic view of the crystal structure of Cs$_{2}$Cu$_{2}$Mo$_{3}$O$_{12}$ (monoclinic, space-group $C2/c$).
}
\label{fig1}
\end{figure}

\begin{figure*}[tbp]
\includegraphics[scale=1]{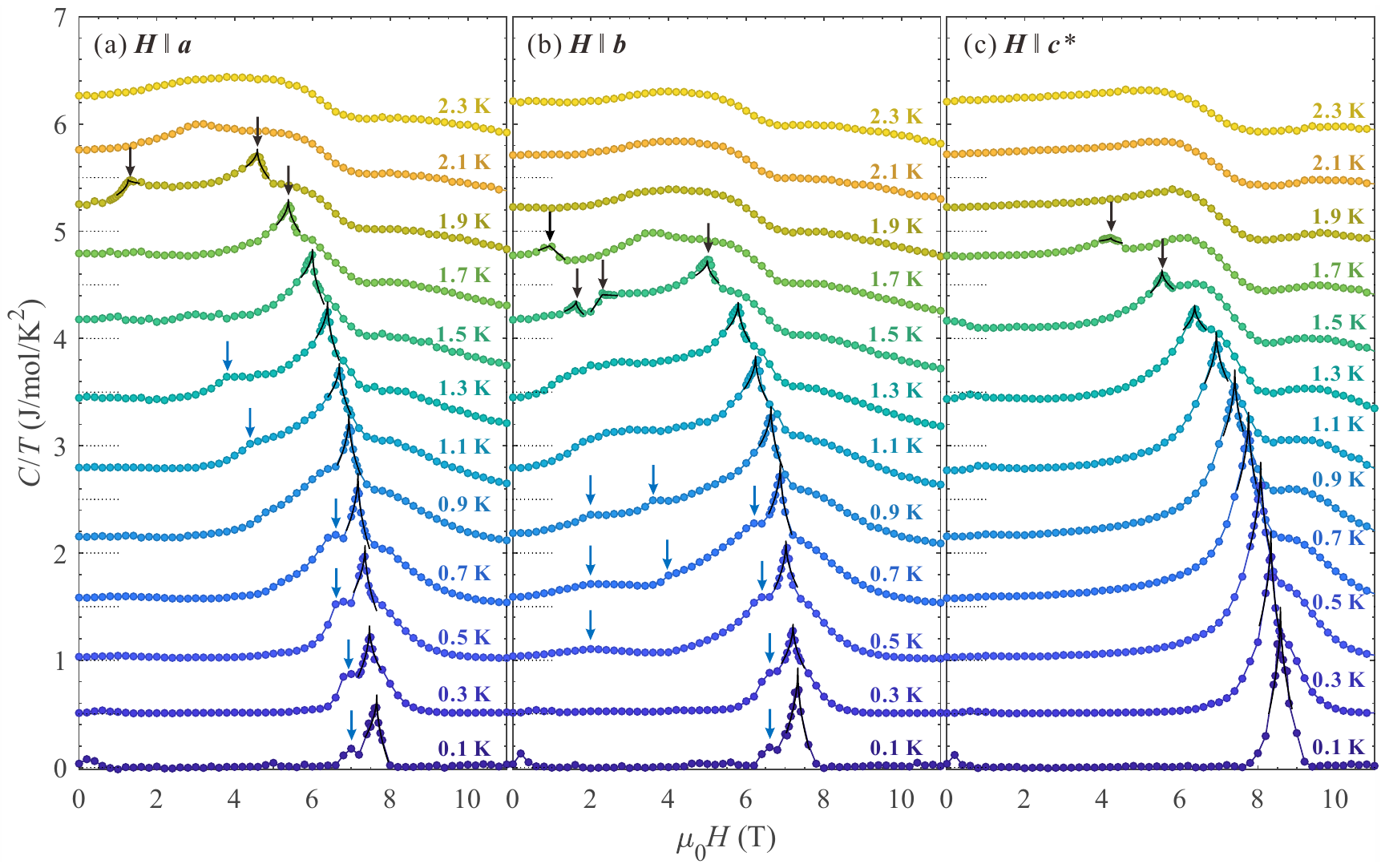}
\caption{Typical heat capacity data for Cs$_{2}$Cu$_{2}$Mo$_{3}$O$_{12}$ in applied magnetic fields along the (a) $a$, (b) $b$, and  (c) $c^{*}$ directions. 
Additional solid curves in the vicinity of sharp peaks and kinks are empirical power-law and linear fits used to pinpoint the transition fields. 
Arrows indicate positions of small features in the data.
For visibility, the scans are offset by $0.5$~J/mol K$^2$ relative to one another.
}
\label{fig2}
\end{figure*}

\section{Experimental details}
A single-crystal sample of Cs$_{2}$Cu$_{2}$Mo$_{3}$O$_{12}$ with mass $91$~$\mu$g was grown by a spontaneous crystallization in a flux method~\cite{Solodovnikov1997}.
Green transparent crystals were obtained as shown in Fig.~\ref{fig1}(a).
The crystal structure [monoclinic $C2/c$, $a=27.932(10)$~{\AA}, $b=5.1134(10)$~{\AA}, $c=20.235(4)$~{\AA}, $\alpha=90^{\circ}$, $\beta=107.124(17)^{\circ}$, $\gamma=90^{\circ}$] was validated using single-crystal x-ray diffraction on a Bruker APEX-II instrument. 
It was found to be totally consistent with the previous report~\cite{Solodovnikov1997}. 
The magnetic properties are due to $S=1/2$ Cu$^{2+}$ cations. 
As illustrated in Fig.~\ref{fig1}(b), CuO$_{4}$ plaquettes form one-dimensional chains along the crystallographic $b$ axis.
The individual chains are paired via MoO$_{4}$ bridges. 
The resulting chain-pairs are separated from one another by the Cs$^{+}$ ions.

All measurements were carried out using the $^{3}$He-$^{4}$He dilution refrigerator insert for the Quantum Design Physical Property Measurement System (PPMS).
The magnetic field was applied along the crystallographic $a$, $b$, and $c^{*}$ directions.
Heat capacity data were collected on a standard Quantum Design relaxation calorimetry option in the range of $0.1$~K~$\leq T \leq2.3$~K and $0$~T~$\leq \mu_{0}H \leq 12$~T.
Magnetization measurements were carried out on a home-made Faraday force magnetometer~\cite{FDM2020}.
Field scans were collected at $0.1$ and $3$~K up to $12$~T.
The absolute values of magnetization were obtained in a calibration measurement at $3$~K using the Vibrating Sample Magnetometer (VSM) option for the PPMS. 

For magnetic torque measurements the crystal was mounted on a CuBe-made cantilever.
For the ${\bm H}\|{\bm a}$ and ${\bm b}$ configurations, we mounted the crystal on a silver-foil sample holder to fix the crystal orientation.
We measured the cantilever deflection by observing a change in electric capacitance $\Delta C$.
The magnetic field was swept continuously from 11 to 0~T, and temperatures set from 0.1 to 2.2~K in 0.1~K steps.
The used sweeping rates were 2.5~Oe/s at 0.1~K, 5~Oe/s at 0.2~K, 10~Oe/s at 0.3 and 0.4~K , and 20~Oe/s for
the rest of the temperature range.
The raw data at each temperature were combined so that the number of the data points is the same for all scans.
The temperature was unstable at low field because sweeping the field heated up the sample holder.
The data whose temperature stability is within $\pm0.05$~K are displayed in the figures.

\section{Results and discussion}
\subsection{Specific heat}
Typical measured field dependent heat capacity scans are shown in Fig.~\ref{fig2}.
Sharp lambda anomalies indicated by solid curves are observed below 2.2 K for all orientations.
These are attributed to a phase transition from the paramagnetic state to the three-dimensional long-range ordered phase (3D LRO).
The obtained critical fields at 0.1~K are anisotropic: $\mu_{0}H_{c}=7.66$~T for the $a$ axis, $7.33$~T for the $b$ axis, and $8.60$~T for the $c^{*}$ axis.
In addition to the sharp lambda peaks, broad but prominent double-hump features are observed near the critical fields for all geometries.
These show the typical behavior of $d=1$, $z=2$ quantum criticality~\cite{Korepin1990,Sachdev1994}, and are observed in other one-dimensional magnets~\cite{Ruegg2008,Breunig2017,Blosser2018} and the isostructural Rb compound~\cite{Hayashida2019}.

\begin{figure}[tb!]
\includegraphics[scale=1]{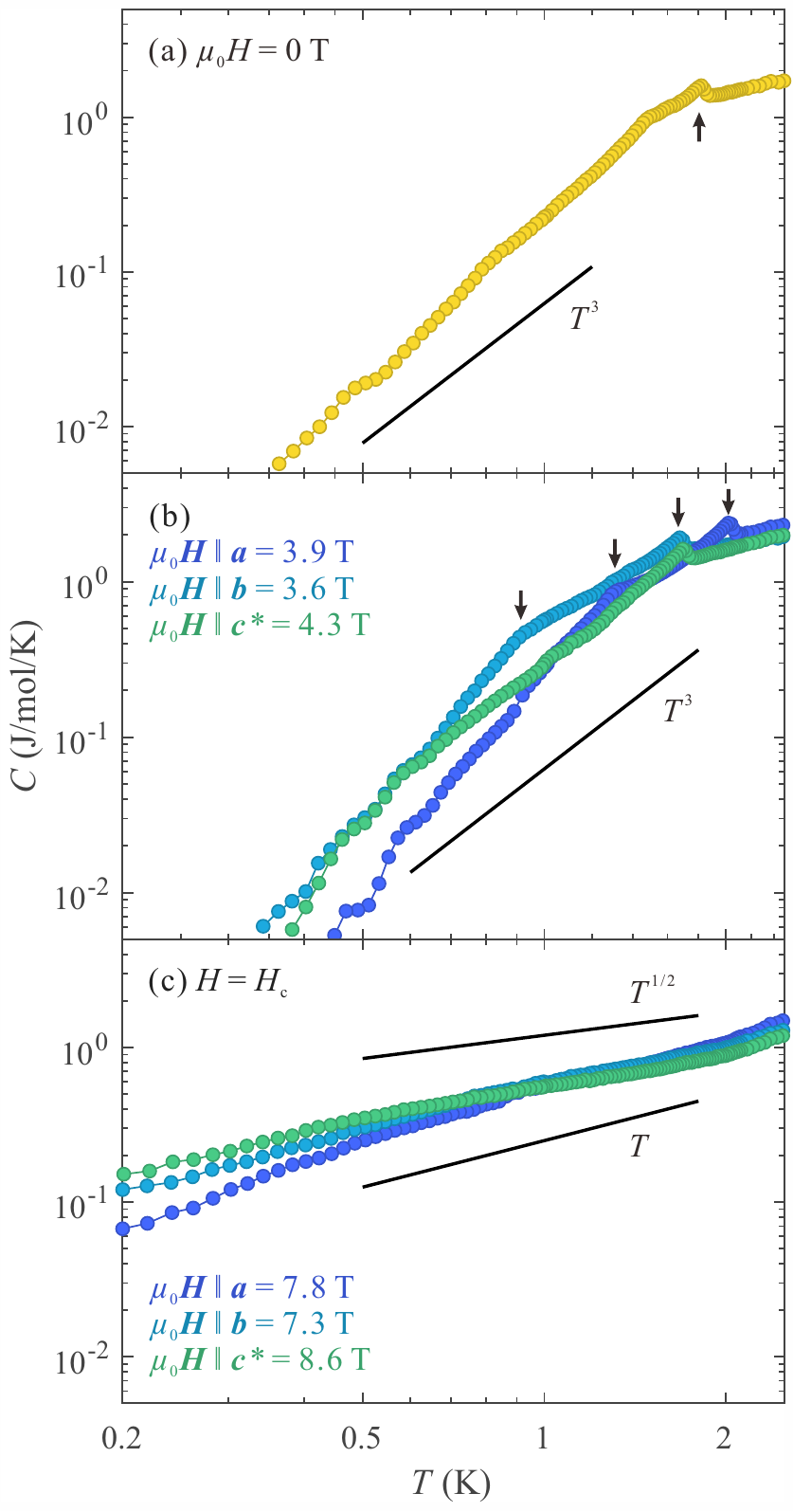}
\caption{Symbols: Temperature scans of the heat capacity measured in  Cs$_{2}$Cu$_{2}$Mo$_{3}$O$_{12}$ at different fields. (a)  Zero applied field.
 (b) Intermediate fields. Arrows indicate lambda-anomalies and kinks. 
 (c) Magnetic fields $H_{c}$ corresponding to quantum phase transitions at full saturation. The solid lines are power laws with different exponents provided for reference.
}
\label{fig3}
\end{figure}

At higher temperatures, the lambda anomalies for each field geometry exhibit qualitatively different behavior.
For ${\bm H}\|{\bm a}$, an additional peak appears at 1.9~K and at 1.4~T [Fig.~\ref{fig2}(a)].
This peak shifts to higher field with increasing temperature, and merges with the other peak at 2.1~K.
For ${\bm H}\|{\bm b}$, double peaks are observed at 1.5~K and around 2~T [Fig.~\ref{fig2}(b)].
With increasing temperature, these peaks shift to lower and higher fields, respectively.
The peak at the high field merges with another lambda peak.
Finally, no additional peak is observed at the boundary for ${\bm H}\|{\bm c}^{*}$ [Fig.~\ref{fig2}(c)].

Characteristic kinks are observed within the domes of the 3D LRO phase as indicated by blue arrows in Fig.~\ref{fig2}.
For ${\bm H}\|{\bm a}$ and ${\bm b}$, small kinks are found beside the sharp lambda-anomalies below 0.8~K.
These indicate the existence of a presaturation phase, as we discuss later in the phase diagram.
Furthermore, we observe temperature-dependent kinks around 4~T for ${\bm H}\|{\bm a}$ and ${\bm b}$, and temperature-independent ones at 2~T for ${\bm H}\|{\bm b}$.
In contrast with the ${\bm H}\|{\bm a}$ and ${\bm b}$ configurations, the specific heat scans in the LRO dome for ${\bm H}\|{\bm c}^{*}$ are rather simple.
Tiny kinks are found in the range of 1.0~K $<T<$ 1.5~K below 1~T, but we suspect that they are experimental artifacts because there is no indication of them in the magnetic torque measurements, as we show later.
Note that small peaks at 0.1~K and 0.2~T for all the geometries are due to nuclear specific heat.

Temperature dependence of the specific heat at zero field is shown in Fig.~\ref{fig3}(a).
Note the logarithmic scale on both axes.
At zero field, the lambda anomaly is clearly observed at $T_{\rm N}=1.82$~K, which is consistent with what is reported in the powder data~\cite{Fujimura2016}.
In the LRO state, the specific heat decays as a power law.
Its thermal dependence is close to $T^{3}$, as expected for three-dimensional spin-waves with a linear gapless dispersion.

In applied fields, across the domes of the 3D LRO phase, constant-$H$ temperature scans of the specific heat show
distinct lambda-anomalies at the phase boundary, as shown in Fig.~\ref{fig3}(b).
In addition, kinks are observed at 1.3~K for the $a$ axis and at 0.9~K for $b$ axis.
The specific heat data rapidly decay below the kinks, suggesting the existence of thermally activated behavior with a gap.
In contrast, the specific heat for ${\bm H}\|{\bm c}^{*}$ follows a $T^{3}$ power law in almost the entire temperature range.
This behavior is attributed to a gapless or slightly gapped magnetic state.
The specific heat at the critical fields is strongly enhanced due to quantum fluctuations, as shown in Fig.~\ref{fig3}(c). 
Since it decays as a power law, a gapless magnetic state is formed.
It is important to note that their thermal behavior at the lowest temperatures does not follow a $T^{1/2}$-law which one would expect for a gapless parabolic dispersion in one dimension.
The observed behavior is distinct from what is observed at the upper critical fields in Rb$_{2}$Cu$_{2}$Mo$_{3}$O$_{12}$~\cite{Hayashida2019}.

\begin{figure}[htbp]
\includegraphics[scale=1]{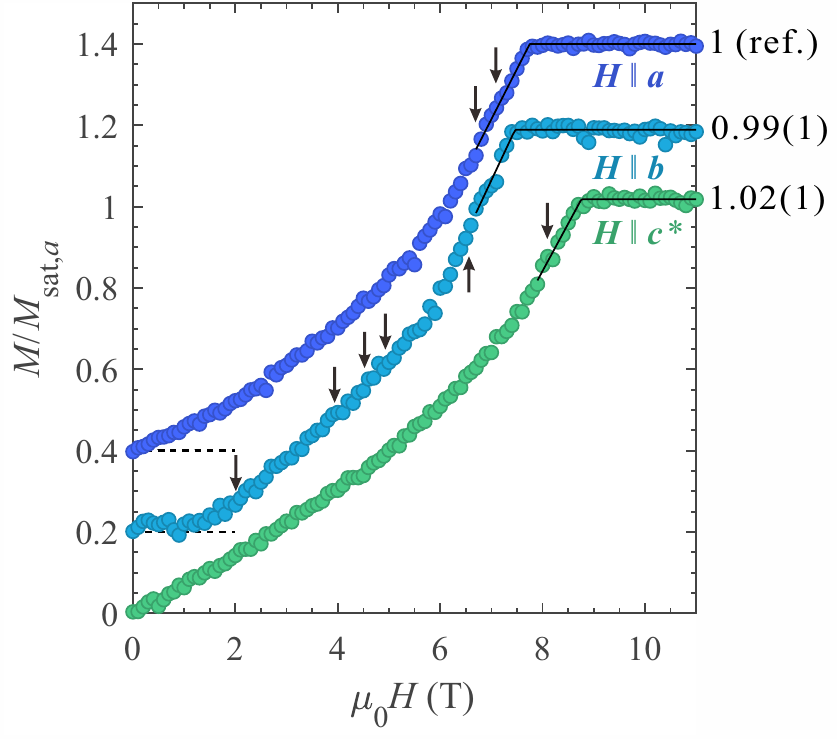}
\caption{Magnetization curves measured in Cs$_{2}$Cu$_{2}$Mo$_{3}$O$_{12}$ at $T = 0.1$~K in fields applied along three crystallographic directions (symbols). 
The data are normalized by the saturation value of 0.84(4)~$\mu_{\rm B}/{\rm Cu}^{2+}$ for the ${\bm H}\|{\bm a}$ configuration.
Solid lines are linear fits to the data in the vicinity of $H_{\rm sat}$.
Arrows indicate the anomalies identified in the torque data.
Two data sets are vertically offset by 0.2 and 0.4 for visibility. 
}
\label{fig4}
\end{figure}

\begin{figure*}[tbp]
\includegraphics[scale=1]{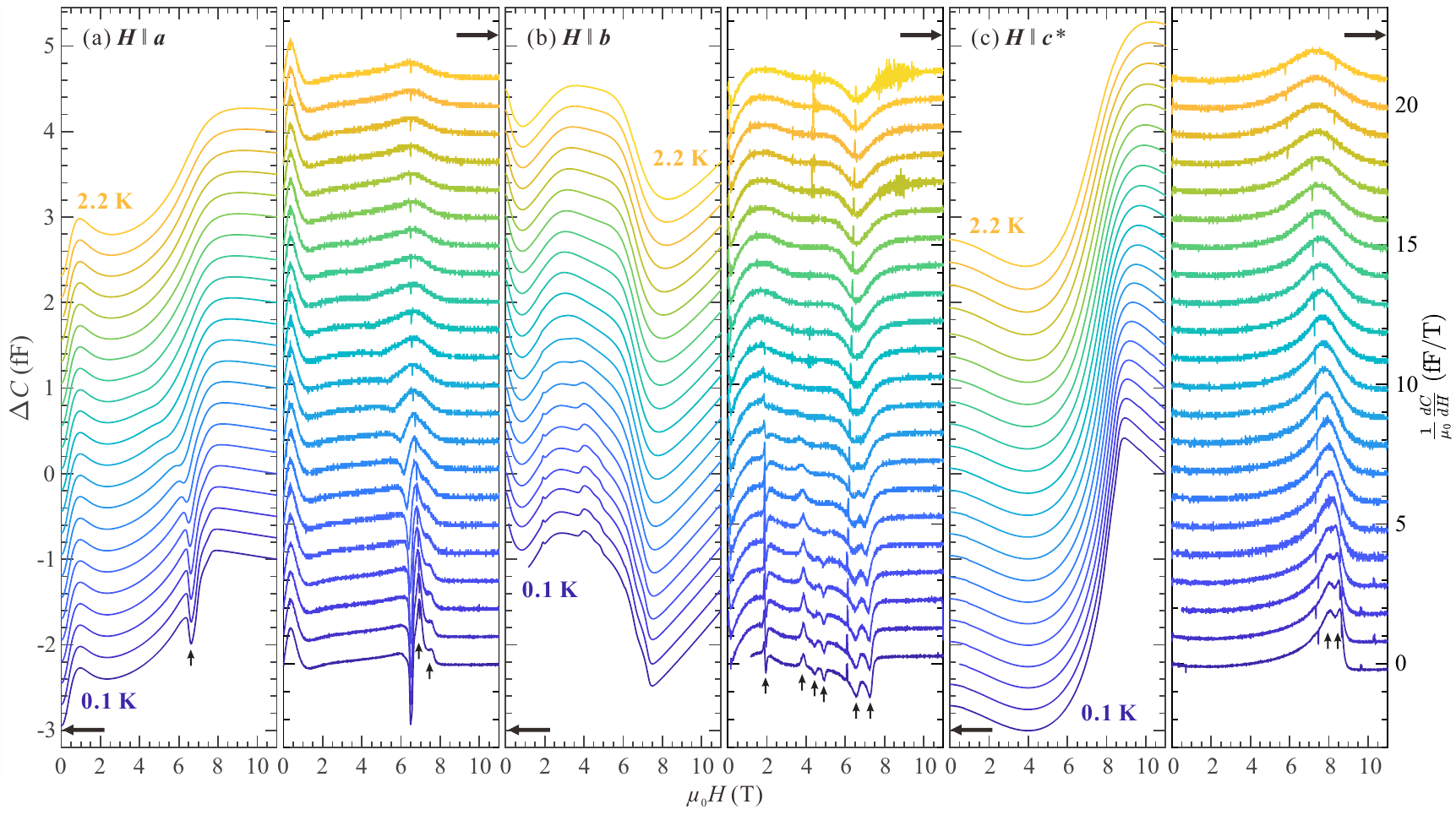}
\caption{$\Delta C$ torque curves (left panel) and their derivative (right panel) for (a) ${\bm H}\|{\bm a}$, (b) ${\bm H}\|{\bm b}$, and (c) ${\bm H}\|{\bm c}^{*}$ configurations.
 The labels of the left and right axes are for $\Delta C$ and $\frac{1}{\mu_{0}}\frac{dC}{dH}$, respectively. 
 Arrows indicate phase transitions as discussed in the text.
 For visibility, the scans are offset by $0.25$~fF for  $\Delta C$ and $1$~fF/T for $\frac{1}{\mu_{0}}\frac{dC}{dH}$ relative to one another.
}
\label{fig5}
\end{figure*}

\subsection{Magnetization}
Magnetization curves for the three field geometries at $T=0.1$~K are shown in Fig.~\ref{fig4}.
The magnetization for the $b$ direction is strongly suppressed at low fields, and a plateau of zero magnetization exists below 2~T.
On the other hand, the magnetization for the $a$ and $c^{*}$ axes increases linearly from zero field.
This behavior is interpreted as a spin-flop reorientation.
It suggests anisotropic exchange interactions,  being the crystallographic $b$ direction the magnetic easy-axis. 

The magnetization at $T=0.1$~K saturates at $\mu_{0}H_{\rm sat}=7.75$~T for the $a$ axis, $7.46$~T for the $b$ axis, and $8.76$~T for the $c^{*}$ axis, in agreement with the critical fields observed in the specific heat data.
In spite of the anisotropic saturation fields, the saturation magnetization for all geometries is very isotropic.
This indicates that the $g$-factor of the Cu$^{2+}$ ion is isotropic even though the spin interactions are anisotropic.
It is also important to note that we see a linear increase of magnetization at saturation. 
This behavior is typical of two- or three-dimensional systems, which is distinct from the square-root singularity found in Rb$_{2}$Cu$_{2}$Mo$_{3}$O$_{12}$~\cite{Hayashida2019} and predicted in general for one-dimensional materials~\cite{Bonner1964}. 
It is likely that exchange interactions between the chains are not negligible. 
On the other hand, the linear approach to saturation may be a feature of the magnetic order in the presaturation phase absent in the Rb compound.

\begin{figure}[tbp]
\includegraphics[scale=1]{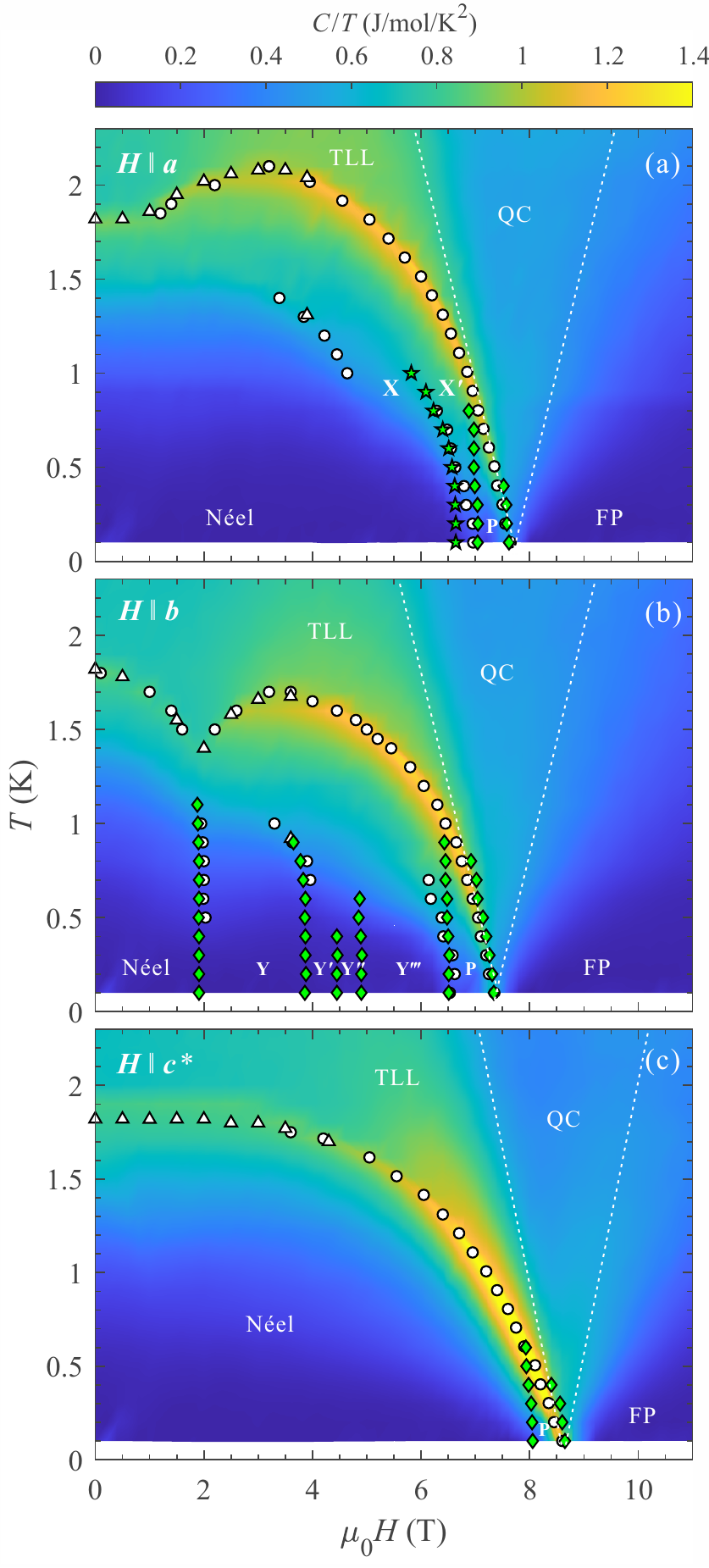}
\caption{Symbols: Magnetic phase diagrams of Cs$_{2}$Cu$_{2}$Mo$_{3}$O$_{12}$ in a magnetic field applied along the (a) $a$, (b) $b$, and (c) $c^{*}$ axes.
White circles and triangles indicate the transitions identified in the field- and temperature scans of the specific heat, respectively.
Green stars and diamonds are the anomalies extracted from the torque data and their derivative, respectively.
The backgrounds show corresponding false color maps of $C(T,H)/T$. The phase regions are labeled as follows: Collinear N{\'e}el order (N{\'e}el);  Presaturation (P); Fully polarized (FP); Tomonaga-Luttinger spin liquid (TLL); and quantum critical regime (QC).
$X$s and  $Y$s indicate prominent but unidentified regions.
Dashed lines are guides for the eye.
}
\label{fig6}
\end{figure}

\subsection{Magnetic torque}

The raw $\Delta C= C(H)-C(11~{\rm T})$ data and their derivative for the three configurations are shown in Fig.~\ref{fig5}.
The magnetic torque curves are greatly dependent on the magnetic field directions and very complicated, evidencing the existence of various magnetic states in the LRO dome.
To pinpoint the phase transitions, we fit our data to empirical distorted Lorentzian functions, as described in Ref.~\citep{Feng2018}.
Note that dips at 1~T for ${\bm H}\|{\bm a}$ and ${\bm b}$ are due to paramagnetic components from the silver-foil sample holder and the CuBe-made cantilever.
Very sharp and temperature-independent features in the derivative of the data are not related to the sample.

First of all, characteristic double peaks are observed at the saturation field in the derivatives for the all three geometries.
While the peaks at higher field correspond to the saturation field, the other ones suggest a transition to a presaturation phase.
Upon increasing temperature, the two peaks merge together, and form one broad peak.
This broad feature probably reflects a cross-over from the partially magnetized to the fully polarized state.

For ${\bm H}\|{\bm a}$, a dip in $\Delta C$ is identified at $6.5$~T as shown in the left panel of Fig.~\ref{fig5}(a).
It shifts to lower fields and its amplitude decreases with increasing temperature.
The most structured curves are observed for ${\bm H}\|{\bm b}$.
We see four additional anomalies at 1.9, 3.8, 4.5, and 4.9~T in the derivative of the data, as indicated by arrows in the right panel of Fig.~\ref{fig5}(b).
The sharp anomaly at 1.9~T is independent of temperature.
We ascribe it to the spin-flop transition observed in magnetization.  
The peak at 3.8~T coincides with the kink observed at 4~T in the specific heat.
However, no signature is found in other measurements that match the dips at 4.5 and 4.9~T.
In contrast, the magnetic torque scans for ${\bm H}\|{\bm c}^{*}$ [Fig.~\ref{fig5}(c)] are rather simple, and no feature is identified except for the presaturation phase.
Note that no anomaly is found below 1~T, where we found the tiny kinks in the range of 1.0~K $<T<$ 1.5~K in specific heat data for ${\bm H}\|{\bm c}^{*}$.

\subsection{Phase diagram}
The magnetic phase diagrams are constructed using the constant-temperature field scans of the specific heat, as shown in Figs.~\ref{fig6}(a) to \ref{fig6}(c) over false-color plots.
The symbols indicate the observed peaks and kinks in specific heat and magnetic torque measurements.
The boundaries of the LRO dome are remarkably anisotropic at high temperatures.
The maximal ordering temperature is visibly enhanced for ${\bm H}\|{\bm a}$, and a bicritical point splitting into two phases is found at 2~T for ${\bm H}\|{\bm b}$.
The phase boundary for ${\bm H}\|{\bm c}^{*}$ is much simpler than for the other directions.

The most striking feature is the emergence of a narrow presaturation phase.
In contrast to the anisotropic behavior at low fields, this phase is independent of the geometry.
This indicates that the presaturation phase is relevant to the physics of the isotropic $J_{1}$-$J_{2}$ spin-$1/2$ chain that predicts a spin-nematic order or a spin-density wave state due to a formation of bound magnon pairs~\cite{Meisner2006,Kecke2007,Vekua2007,Hikihara2008,Sudan2009,Meisner2009,Zhitomirsky2010,Sato2013,Nishimoto2015}.
It is noted that the NMR study found the spin-nematic TLL state above 7.3~T in the paramagnetic phase~\cite{Hoshino2014,Goto2017}. 
Therefore, it is suggested that a spin nematic order is developed in the presaturation phase in Cs$_{2}$Cu$_{2}$Mo$_{3}$O$_{12}$. 

Another unusual feature is the existence of complex magnetic transitions in the intermediate fields for ${\bm H}\|{\bm b}$ as shown in Fig.~\ref{fig6}(b).
First, the spin-flop transition at 2~T splits the LRO dome into two regions.
Due to the easy-axis anisotropy and the frustration, it is likely that the transition occurs from a collinear N{\'e}el state to an incommensurate structure.
In addition, it is noteworthy that multiple transitions are found in the high field phase.
Up to five different regions are identified at low temperatures.
Since we do not see any indications in the magnetization for these transitions (see Fig.~\ref{fig4}), they may be attributed to the transverse component of the spin. 
On the contrary, the phase diagrams for the other geometries are simpler than that for ${\bm H}\|{\bm b}$.
For ${\bm H}\|{\bm a}$, two transitions are identified at 4 and 6.5~T below the presaturation phase [Fig.~\ref{fig6}(a)].
There is no clear indication of an additional transition for ${\bm H}\|{\bm c}^{*}$ [Fig.~\ref{fig6}(c)].

We compare the measured phase diagrams to the ones in other FM-AFM spin chain materials.
Spin-density wave states~\cite{Masuda2011,Mourigal2012,Nawa2013,Willenberg2016,Heinze2019,Nawa2017,Asai2020,Pregelj2019} and presaturation phases suggesting a spin nematic order~\cite{Buttgen2014,Orlova2017,Gen2019,Willenberg2012,Schapers2013,Willenberg2016,Feng2018,Heinze2019,Pregelj2015,Weickert2016,Pregelj2019} have been previously found.
The observed presaturation phase for Cs$_{2}$Cu$_{2}$Mo$_{3}$O$_{12}$ exists in a narrow range between $\sim0.9M_{\rm sat}$ and $M_{\rm sat}$, which approximately coincides with those ranges for LiCuVO$_{4}$~\cite{Gen2019} and $\beta$-TeVO$_{4}$~\cite{Pregelj2019}.
The most significant difference is that the collinear N{\'e}el state is formed at zero field in Cs$_{2}$Cu$_{2}$Mo$_{3}$O$_{12}$, whereas a spin spiral order was found in the other candidate materials~\cite{Masuda2011,Mourigal2012,Nawa2013,Willenberg2012,Willenberg2016,Heinze2019,Nawa2017,Asai2020,Pregelj2015,Pregelj2019}.
Thus, Cs$_{2}$Cu$_{2}$Mo$_{3}$O$_{12}$ is a rare FM-AFM spin chain material with an easy-axis anisotropy.

A FM-AFM spin chain with an easy-axis anisotropy has been less studied than that with an easy-plane anisotropy~\cite{Igarashi1989,Tonegawa1990,Meisner2009}.
In the Ising limit of the anisotropy, fully polarized ferromagnetic and up-up-down-down antiferromagnetic phases appear below and above $J_{2}/|J_{1}|=1/2$, respectively.
When a weak easy-plane anisotropy is introduced, a partially polarized ferromagnetic phase is formed between the fully polarized ferromagnetic and up-up-down-down phases~\cite{Igarashi1989,Tonegawa1990}.
In addition, it has been proven that an easy-axis anisotropy enhances a multimagnon bound state~\cite{Meisner2009}.
However, the detailed phase diagram with a weak easy-axis anisotropy, which is the case in the present result, remains unclear.
Further experimental and theoretical investigation of the ground state and spin model is crucial to elucidate  the presaturation phase and  multiple transitions in Cs$_{2}$Cu$_{2}$Mo$_{3}$O$_{12}$.

\section{Conclusion}
Magnetic and thermodynamic measurements using a single-crystal sample provide several important findings for the nature of the quantum magnetism in Cs$_{2}$Cu$_{2}$Mo$_{3}$O$_{12}$.
(i) The $T^3$-like behavior of the specific heat in zero field as well as the linear magnetization near saturation may indicate that inter-chain interactions are more important than in the Rb-based system.
(ii) The observed spin-flop transition indicates easy-$b$-axis magnetic anisotropy and a collinear magnetic structure at zero field. The anisotropy is that of interactions, since the saturation magnetization is remarkably isotropic.
(iii) A novel presaturation phase is found for all geometries. We speculate that it represents spin-nematic order or other multi-magnon condensate. 
(iv) In the LRO dome, multiple transitions are identified applying the field along the $b$ direction. 
Further work on the spin model is necessary to elucidate the unique magnetic states in Cs$_{2}$Cu$_{2}$Mo$_{3}$O$_{12}$.

\section*{Acknowledgements}
This work was supported by Swiss National Science Foundation under Division II.

%

\end{document}